\documentclass[12pt,draftclsnofoot, onecolumn]{IEEEtran}
\usepackage{amsmath, amsthm, amssymb} 
\usepackage{graphicx}
\usepackage{epstopdf}
\usepackage[noadjust]{cite}

\allowdisplaybreaks
\ifCLASSINFOpdf

\else
  
\fi

\begin{document}

\title{Low-Complexity Hybrid Beamforming for Massive MIMO Systems in Frequency-Selective Channels}
\author{Sohail Payami, Mathini Sellathurai, Konstantinos Nikitopoulos

\thanks{Manuscript submitted February 04, 2019; accepted
March 29, 2018. This work was supported by the U.K. Engineering and
Physical Sciences Research Council under Grant EP/M014126/1, Large Scale
Antenna Systems Made Practical: Advanced Signal Processing for Compact
Deployments. The majority of the work has been carried out with the first author being with Heriot-Watt University, UK. This work is also partially supported by the 5G Innovation Centre, University of Surrey and the National Physics Laboratory (NPL), UK.}% <-this % stops a space

\thanks{S.Payami and K. Nikitopoulos are with Institute for Communication Systems, 5G Innovation Centre,  Wireless Systems Lab, University of Surrey, Guildford GU2 7XH, United Kingdom. (e-mails: $\lbrace$sohail.payami, k.nikitopoulos$\rbrace$surrey.ac.uk)}  
\thanks{M. Sellathurai (Corresponding author) is with Heriot-Watt University, Edinburgh EH14 4AS, U.K. (email: m.sellathurai@hw.ac.uk)}}

\maketitle


\ifCLASSOPTIONcaptionsoff
  \newpage
\fi

\begin{abstract}

Hybrid beamforming for frequency-selective channels is a challenging problem as the phase shifters provide the same phase shift to all of the subcarriers. The existing approaches solely rely on the channel's frequency response and the hybrid beamformers maximize the average spectral efficiency over the whole frequency band. Compared to state-of-the-art, we show that substantial sum-rate gains can be achieved, both for rich and sparse scattering channels, by jointly exploiting the frequency and time domain characteristics of the massive multiple-input multiple-output (MIMO) channels. In our proposed approach, the radio frequency (RF) beamformer coherently combines the received symbols in the time domain and, thus, it concentrates signal's power on a specific time sample. As a result, the RF beamformer flattens the frequency response of the ``effective'' transmission channel and reduces its root mean square delay spread. Then, a baseband combiner mitigates the residual interference in the frequency domain. We present the closed-form expressions of the proposed beamformer and its performance by leveraging the favorable propagation condition of massive MIMO channels and we prove that our proposed scheme can achieve the performance of fully-digital zero-forcing when number of employed phase shifter networks is twice the resolvable multipath components in the time domain.

\end{abstract}

\section{Introduction}
Fully-digital massive multiple-input multiple-output (MIMO) systems are considered as one of the key technologies to scale up the data rates in cellular communications \cite{6375940,6798744,6206345}. Such structures require a dedicated radio frequency (RF) chain per antenna which makes fully-digital beamforming an expensive and a power hungry technology \cite{8030501,7389996,6717211}. To overcome these issues, hybrid analog-and-digital beamformers have been considered as an alternative solution to fully-digital systems in massive MIMO scenarios \cite{6717211}. In hybrid structures, a small number of RF chains are connected to a large number of antennas through a network of low-cost phase shifters \cite{8371237}. The design of hybrid beamformers is a challenging task as it requires solving a difficult nonconvex optimization problem due to the nonconvex constraints that are imposed by the phase shifters. Compared to fully-digital systems, it has been shown that hybrid beamformers can provide a significantly higher energy efficiency \cite{8382230,7539303,7961162} and a competitive spectral efficiency \cite{8295113,Globcome2018, 7542170} when frequency-flat channel models are considered. Although there are many papers that have investigated hybrid beamforming for frequency-flat channels such as \cite{8371237,7433949} and the references therein; there is a limited work on the design of such beamformers for frequency-selective channels.

Designing hybrid beamformers for frequency-selective channels is a more challenging problem as the RF beamformer applies the same phase shift to the whole frequency band. State-of-the-art papers on hybrid beamforming, such as \cite{7504275,8323164,7448873,7248579, 7588162,8030501, 7880698,7913599,7472284,7454701,7925850} and references therein, generally exploit the sparsity of millimeter wave (mmWave) channels and employ various optimization tools to calculate the beamforming weights. However, it is not possible to deduce a closed-form expression or even an approximation of the beamformer or its performance. Such channels have special properties as they consist of only a few multipath components which enables development of hybrid beamformers for low-ranked channel matrices \cite{7472284,7913599,7887656}. In \cite{8323164,7448873,7248579, 7588162,8030501, 7880698,7913599,7472284,7454701,7925850} the authors exploit the sparse nature of the mmWave channel and employ compressive sensing methods such as distributed compressive sensing \cite{7454701}, projected gradient and alternating minimization methods \cite{8323164}, orthogonal matching pursuit and gradient pursuit algorithms  \cite{8323164,7588162}. Recently, it has been shown that a modified form of narrowband hybrid beamforming techniques can be applied to the RF beamformer for mmWave frequency-selective scenarios \cite{7925850,7913599,7887656}. In \cite{7925850,7913599}, first the average of the channel matrices over different subcarriers is calculated, and then the narrowband beamformers of \cite{7389996} or \cite{7925850} are applied to the average channel. In short, hybrid beamforming for frequency-selective channels is a relatively new area and many challenges need to be addressed ranging from efficient channel estimation to the design of the corresponding beamformer. Even when the perfect channel state information (CSI) is available, there are many unanswered questions on the the design of the beamformer, such as:
\begin{itemize}
\item Hybrid beamformers are suitable for massive MIMO scenarios when a large number of antennas are required. Instead of using complex optimization tools to evaluate the performance of the beamformers, is it possible to exploit the statistical properties of massive MIMO and benefit from the deterministic and favorable conditions in such scenarios?
\item Is it possible to derive closed-form approximations of the beamformer and its performance so that it can be used as a design guide?
\item If extra phase shifters are available, while the number of antennas and RF chains remain constant, how can such phase shifters be exploited? To the best of authors' knowledge the hybrid beamforming designs in the literature can only support a specific number of phase shifters and they are not scalable; i.e. if extra phase shifters are available then it is not clear how they can be exploited. 
\item How many phase shifters are required to achieve the performance of a fully-digital system in frequency-selective channels? For example in \cite{MOlischAntennaselection2005Journal,7542170}, it is shown that the exact performance of a fully-digital beamformers in frequency-flat channels can be achieved if the number of the RF chains is twice larger than the number of the transmit streams. However, it is not clear that how many phase shifters are required to achieve the performance of fully-digital beamforming in frequency-selective channels. 
\end{itemize} 
This paper answers all the above questions by exploiting the characteristics of massive MIMO channels both in the time and frequency domains. Traditionally, hybrid beamformers are designed in the frequency domain such that the average spectral efficiency over all of the subcarriers is maximized \cite{8323164,7504275,7448873,7248579, 7588162,8030501, 7880698,7913599,7472284,7454701,7925850}. However, we view the hybrid beamforming problem as a two-stage beamformer where the RF beamformer is designed by accounting for the impulse response of the channel. Then, the baseband combiner is designed according to the frequency response of the resulting effective channel which includes the impact of the RF beamformer on the propagation channel. In this direction, the contributions of this paper are summarized as:
\begin{itemize}
\item We propose a low-complexity technique for hybrid beamforming for frequency-selective channels. In this approach, the RF beamformer coherently combines the received samples from different time instants such that the energy of the desired symbol is focused onto an specific time sample. As a result, the large dimensional frequency-selective propagation channel is converted to an effective channel which has smaller dimensions, smaller root mean square (RMS) delay spread, and more flat frequency response. By leveraging the favorable propagation in massive MIMO systems, we derive closed-form expressions of the asymptotic achievable sum-rate by the RF beamformer and the capacity of the effective channel. Our results indicate that the proposed RF beamformer provides a promising sum-rate in the low signal-to-noise ratio (SNR) regime; however, its performance saturates at high SNRs. To overcome this limitation, the baseband combiner is designed to compensate for the residual interference in the frequency-domain, and thus, to enhance the provided performance also at the high SNR regime. Our method not only allows for the derivation of simple and tractable closed-form expressions of the beamformer and its achievable sum-rates, but also it provides a better performance in both rich and sparse scattering channels compared to state-of-the-art.
\item We investigate the RMS delay spread behavior of the resulting effective channel. The proposed technique substantially reduces the RMS delay spread of the effective channel as the number of the antennas grows large. This is also equivalent to an effective channel which converges to a frequency-flat channel when the number of the antennas goes large. Based on this behavior, the asymptotic capacity of the effective channel is derived when the number of the antennas goes large.
\item We propose a new structure to address the beamformer design when extra phase shifter networks and delay lines are available. As an extension to \cite{MOlischAntennaselection2005Journal,7542170} that focus on hybrid beamforming over frequency-flat channels, we also derive the number of the required phase shifters to achieve the performance of fully-digital zero-forcing in frequency-selective channels. In particular, we prove that the exact performance of fully-digital beamformer can be archived when the number of the phase shifter networks is twice the number of the resolvable multipath components in the channel impulse response.
\end{itemize}

\begin{figure*}[htb!]
\centering
\includegraphics[width=\textwidth]{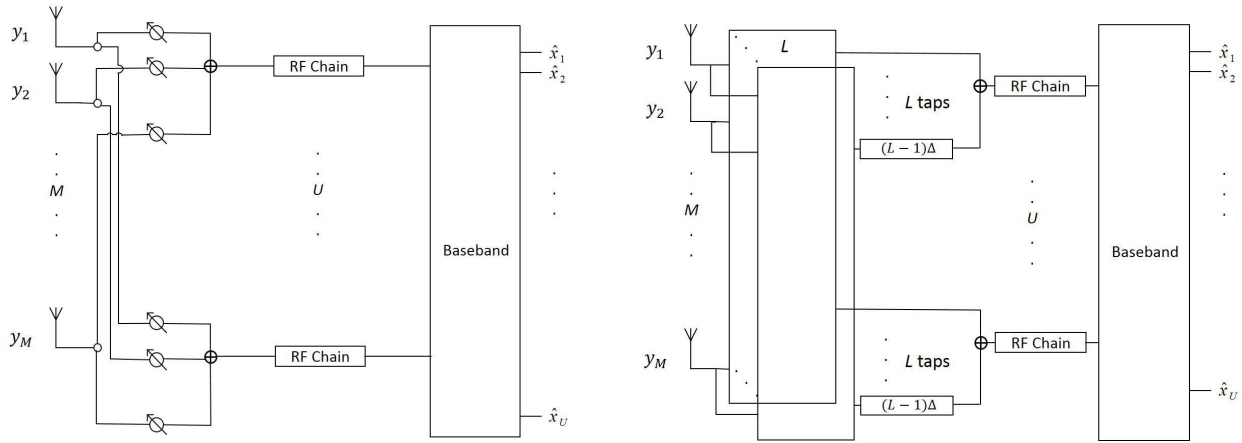}
\caption{Block diagram of 1-tap (left) and $L$-tap (right) hybrid beamformers.}
\label{Blocks}
\end{figure*}

\subsection{Notations} 
The following notation is used throughout this paper: $\mathbb{R}$ and $\mathbb{C}$ are the field of real and complex numbers. $\mathbf{A}$ and $\mathbf{a}$ represent a matrix and vector. $\mathbf{a}_m$ is the $m$th column of $\mathbf{A}$. $A_{mn}$ and $\vert A_{mn}\vert$ denote the $(m,n)$ element of $\mathbf{A}$ and its magnitude. $\mathbf{A}^{-1}$, det$( \mathbf{A} )$, $\mathbf{A}^\text{T}$ and $\mathbf{A}^\text{H}$ denote inverse, determinant, transpose and Hermitian of $\mathbf{A}$, respectively. $\mathcal{CN}(\mathbf{a},\mathbf{A})$ presents a random vector of complex Gaussian distributed elements with expected value $\mathbf{a}$ and covariance matrix $\mathbf{A}$. Finally, $\text{E}_M[ a]$ and $\text{Var}_M (a)$ denote the expected value and variance of $a$ with respect to $M$.

 \section{System Model}
Consider the uplink of a single-cell massive MIMO system where $U$ single-antenna users at time index $n$ transmit the signal vector $\mathbf{x} (n) \in \mathbb{C}^{U \times 1}$ to the base station with $M\gg U$ antennas. The elements of $\mathbf{x}(n)$ are independent and identically distributed (i.i.d.) with E$[\mathbf{x}(n) \mathbf{x}^\text{H}(n)]= P_\text{t} \mathbf{I}_U$ where $P_\text{t}$ is the transmit power of each user. The impulse response of the wireless channel matrix $\mathbf{H}(\tau)$ is described as
\begin{equation}
\label{eq:timevaryingChannel1}
\mathbf{H}(\tau)=\sum_{l=0}^{L-1} \mathbf{H}_{l} \delta(\tau-\tau_l),
\end{equation}
where $L$, $\delta (\tau)$ and $\mathbf{H}_{l} \in \mathbb{C}^{M \times U}$ denote the total number of the delay bins, Dirac delta function and the channel matrix at the $l$-th delay bin, respectively. It is noted that the delay resolution of the system is $\Delta=\tau_l-\tau_{l-1}=1/B$ where $B$ presents the signal bandwidth \cite{6666103}. The discrete impulse response of the channel impulse response in equation (\ref{eq:timevaryingChannel1}) can be written as 
\begin{equation}
\mathbf{H}(\tau)=\mathbf{H}(n\Delta)=\sum_{l=0}^{L-1} \mathbf{H}_{l} \delta(n\Delta-l\Delta).
\end{equation}
The channel matrix $\mathbf{H}_{l}$ consists of slow and fast fading parameters which are denoted by matrix $\mathbf{D}_l \in \mathbb{R}^{U \times U}$ and $\mathbf{H}_{\text{w}l} \in \mathbb{C}^{M \times U}$, respectively. Hence, the $L$-tap frequency-selective channel matrix $\mathbf{H}$ is reformulated as 
\begin{equation}
\label{eq:timevaryingChannel}
\mathbf{H}(n)=\sum_{l=0}^{L-1} \mathbf{H}_{l} \delta(n-l)=\sum_{l=0}^{L-1} \mathbf{H}_{\text{w}l}  \mathbf{D}_l^{1/2} \delta(n-l).
\end{equation}
We assume that the nonzero elements of the diagonal matrix $\mathbf{D}_l$ are denoted by $d_{lu}$ and modeled as \cite{6189014}
\begin{equation}
\label{d_lk}
d_{lu}=\text{exp}(-\psi_u l)/ \sum_{l'=0}^{L-1} \text{exp}(-\psi_u l'),
\end{equation}
where $\psi_u=(u-1)/5$, $\forall u\in \lbrace 1,...,U \rbrace$ and $\sum_{l=0}^{L-1} d_{lu}=1$. Moreover, the distribution of the elements of $\mathbf{H}_{\text{w}l}$ follow $\mathcal{CN}(0,1)$ and they are i.i.d. and uncorrelated. The relationship between $\mathbf{x}(n)$ and the received signal vector $\mathbf{y}(n) \in \mathbb{C}^{M \times 1}$ is 
\begin{equation}
\mathbf{y}(n)= \sum_{l=0}^{L-1} \mathbf{H}(l)\mathbf{x}(n-l) +\mathbf{z}(n),
\end{equation}
where $\mathbf{z} \in \mathbb{C}^{M\times 1}$ denotes the i.i.d. zero-mean additive white Gaussian noise vector with variance $\sigma_\text{z}^2$ and E$[\mathbf{z}(n) \mathbf{z}^\text{H}(n)] = \sigma_\text{z}^2 \mathbf{I}_M$. The base station is assumed to have perfect knowledge of the CSI and employs the combiner matrix in the form of 
\begin{equation}
\label{CombinerGenral}
\mathbf{W} (n)= \sum_{l=-L+1}^{0} \mathbf{W}_{l} \delta(n-l),
\end{equation} 
where $\mathbf{W}_{l} \in \mathbb{C}^{U \times M}$. Further discussion on the combiner design will be provided later. In hybrid beamforming, the combiner matrix $\mathbf{W}$ consists of a baseband combiner $\mathbf{W}_\text{BB}  \in \mathbb{C}^{U \times U}$ and an RF beamformer $\mathbf{W}_\text{RF}\in \mathbb{C}^{U \times M}$. As shown in Fig \ref{Blocks}, two RF beamforming structures are considered in this paper which we refer to them as 1-tap and $L$-tap beamformers. The 1-tap beamformer is the traditional fully-connected structure where there is a connection from each RF chain to all of the antennas via a phase shifter and an adder. In this approach, the phase shifter network is placed on an IC. In this paper, we will also investigate the impacts of having extra phase shifters. The $L$-tap beamformer can be viewed as having $L$ ICs and delay lines $l\Delta$, $l\in \lbrace 1,...,L-1 \rbrace$ which their outputs are connected via an adder. Since the 1-tap beamformer is an special case of the $L$-tap method, we use a generic notation to represent the elements of the RF beamforming matrix $\mathbf{W}_{\text{RF},l}$ as
\begin{equation} 
W_{\text{RF},lum}= 1/\sqrt{M}\text{e}^{j \theta_{lum}},
\end{equation} 
where $ \: \theta_{lum} \in[0,2\pi),  \: l \in \lbrace 0,\:...,\: L-1\rbrace,  \:  u \in \lbrace 1,\:...,\: U\rbrace$ and  $ m \in \lbrace 1,\:...,\: M\rbrace$. Considering the impulse response of the combiner in (\ref{CombinerGenral}), the impulse response of the $L$-tap RF beamformer in Fig. 1 is in the form of
\begin{equation}  
\mathbf{W}_{\text{RF}}(n)=\sum_{l=-L+1}^0 \mathbf{W}_{\text{RF},l} \delta(n-l).
\end{equation}

Throughout this paper, we will frequently refer to effective channel matrix $\mathbf{H}_\text{e} \in \mathbb{C}^{U \times U}$ and the effective noise vector $\mathbf{z}_\text{e}(n)  \in \mathbb{C}^{U \times 1}$ which for the $L$-tap beamformer they are defined as 
\begin{equation}
\label{ImpulseResponsesLtap}
\begin{cases}
\mathbf{H}_\text{e, $L$-tap}(n)=\sum_{l=-L+1}^{l=L-1} \mathbf{W}_{n-l}  \mathbf{H}_l ,\\
\mathbf{z}_\text{e, $L$-tap}(n) =\sum_{l=-L+1}^{0} \mathbf{W}_{n-l}  \mathbf{z}_l .
\end{cases}
\end{equation}
For the 1-tap beamformer the effective channel matrix and noise vector are
\begin{equation}
\label{ImpulseResponses1tap}
\begin{cases}
\mathbf{H}_{\text{e, $1$-tap}(n)}= \mathbf{W}_{0}  \mathbf{H}_l\delta(n),\\
\mathbf{z}_\text{e, $1$-tap}(n) = \mathbf{W}_{0}  \mathbf{z}_l\delta(n).
\end{cases}
\end{equation}
In order to present our hybrid beamformer and its performance, we will firstly review matched filtering (MF) and ZF in the fully-digital systems.

\subsection{Background}
\begin{figure}
    \centering
          \includegraphics[width=.5\textwidth]{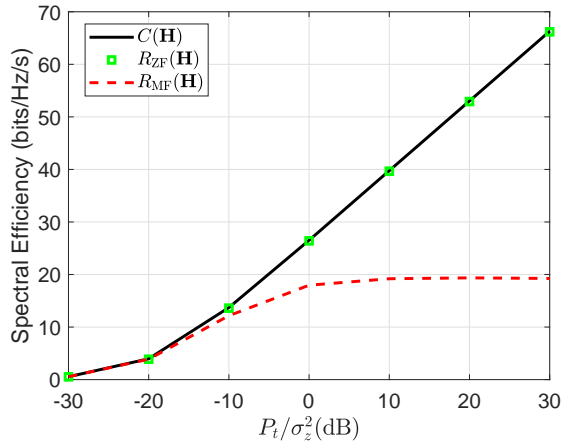}
        \caption{Spectral efficiency v.s. $P_\text{t}/\sigma_\text{z}^2$. The simulation parameters are $K=128$, $M=100$, $U=4$ and $L=4$.}
        \label{fig:Fully-digital}
        \vspace{-2mm}
\end{figure}

The capacity of frequency-selective channel $\mathbf{H}(n)$ is expressed as \cite{983319}
\begin{equation}
\label{eq:fsCapacity}
C(\mathbf{H})=\frac{1}{K}\sum_{k=1}^{K}  \log_2 \text{det}\Big( \mathbf{I} + \rho \tilde{\mathbf{H}}^\text{H} (k) \tilde{\mathbf{H}} (k)) \Big),
\end{equation}
where $\rho=P_\text{t}/\sigma^2_{z}$ is a measure of SNR and 
\begin{equation}
\label{FFTeq}
\tilde{\mathbf{H}}(k)=\sum_{l=0}^{L-1} \mathbf{H}_l \text{exp}(-\frac{j2\pi l k}{K}),
\end{equation}
is the frequency response of the channel at subcarrier $k\in \lbrace1,2,...,K \rbrace$. In massive MIMO systems, linear beamformers such as MF and ZF provide a near-optimal performance \cite{6375940}. MF can be performed either in the frequency domain by multiplying the transfer function of the received signal with $\tilde{\mathbf{H}}^\text{H}(k)$; or in the time domain \cite{6189014}, i.e. convolution with 
\begin{equation}
\label{MFDigi}
\mathbf{W}_\text{MF}(n)=\mathbf{H}^\text{H}(-n)/\sqrt{M}=\sum_{l=-L+1}^0 \mathbf{H}^\text{H}_{(-l)} \delta(n-l)/\sqrt{M}.
\end{equation} 
It is noted that $1/\sqrt{M}$ in (\ref{MFDigi}) is simply a normalization factor and it does not change the overall performance. As an example, for the time domain implmenetaion, cosider a two-tap channel where $\mathbf{H}(n)=\sum_{l=0}^{1} \mathbf{H}_{l} \delta(n-l)$. Then, the impulse response of MF is 
\begin{equation}
\mathbf{W}_\text{MF}(n)=\frac{1}{\sqrt{M}}\bigg(\mathbf{H}^\text{H}_{0} \delta(n)+\mathbf{H}^\text{H}_{1} \delta(n+1)\bigg).
\end{equation}
MF in massive MIMO scenarios achieves a near-optimal spectral efficiency at low SNRs but its performance saturates at high SNRs due to inter-user and inter-symbol interference \cite{6189014}. Moreover, MF reduces the RMS delay spread of each user's effective channel with $1/\sqrt{M}$ \cite{6666103}. On the other hand, ZF with 
\begin{equation}
\label{ZFDigi}
\tilde{\mathbf{W}}_\text{ZF}(k)=\big (\tilde{\mathbf{H}}^\text{H}(k) \tilde{\mathbf{H}}(k) \big)^{-1} \tilde{\mathbf{H}}^\text{H}(k) ,
\end{equation} 
provides a near-optimal performance when $U \ll M$. Equation (\ref{ZFDigi}) implies that the pseudo-inverse operation can be viewed as a two-stage beamformer where the first-stage MF is followed by a second-stage ZF, i.e. matrix inversion over the $U \times U$ dimensional effective channel. To summarize this subsection, Fig. \ref{fig:Fully-digital} presents a comparison between the channel capacity $C(\mathbf{H})$ and the sum-rates $R_\text{ZF}$ and $R_\text{MF}$ by ZF and MF, respectively. The simulations are averaged over 1000 Monte-Carlo realizations for $M=100$, $U=4$, $L=4$ and $K=128$.

\subsection{Overview of the Existing Approaches}
According to the conventional approach in the literature, the objective function for designing a 1-tap beamformer in frequency-selective channels is \cite{7448873}
\begin{equation}
\underset{\mathbf{W}_\text{B} (k), \mathbf{W}_\text{RF}}{\arg\max} R= \frac{1}{K}\sum_{k=1}^{K} R(k),
\end{equation}
where $R(k)$ is the sum-rate at the $k$-th subcarrier; and it is expressed as
\begin{align}
\label{1tapOptimizationProblem}
R(k)=  \log_2 \text{det}\Big(& \mathbf{I} + \rho \big( \tilde{\mathbf{W}}_\text{B} (k) \mathbf{W}_\text{RF}   \mathbf{W}_\text{RF}^\text{H}   \tilde{\mathbf{W}}_\text{B}^\text{H} (k) \big)^{-1} \times \\ 
&\tilde{\mathbf{W}}_\text{B} (k)   \mathbf{W}_\text{RF} \tilde{\mathbf{H}} (k)   \tilde{\mathbf{H}}^\text{H} (k)   \mathbf{W}_\text{RF}^\text{H}  \tilde{\mathbf{W}}_\text{B}^\text{H} (k) \Big).\nonumber
\end{align}
The design criteria in (\ref{1tapOptimizationProblem}) indicates that RF beamformer should be designed such that it maximizes the average spectral efficiency over all of the subcarriers. In this direction, various optimization methods are proposed in the literature to solve this problem considering the nonconvex modulus constraint that is imposed by the phase shifters. In other words, state-of-the-art methods, such as \cite{7913599,7448873,8306126,8323164} and references therein, design the hybrid beamformer according to the frequency response of the channel. 

Based on the approaches in state-of-the-art \cite{7913599,7448873,8306126,8323164} and many references therein, 
\begin{itemize}
  \item It is not possible to derive a closed-form approximation of the performance.
  \item The computational complexity becomes relatively high.
\item The extension to rich scattering channels may not possible as in \cite{7448873,8306126,8323164}.
\item If extra phase shifters are available then it is not clear how the additional phase shifters could be exploited.
\item It is not clear how many phase shifters are required to achieve the performance of fully-digital beamforming?
\end{itemize}
By revisiting the approach towards the design of hybrid beamformers in frequency-selective channels, we will show that the these challenges can be addressed in massive MIMO scenarios.

\section{Hybrid Analog-and-Digital Beamforming}
In this paper, we rely on the deterministic behaviors of massive MIMO both in time and frequency domains to propose a low-complexity hybrid beamforming technique for the 1-tap and $L$-tap structures over frequency-selective channels. In addition to the closed-form expressions of the beamformer, we present asymptotic expressions of the performance in terms of sum-rate. Moreover, the RMS delay spread of such systems will be studied.  In the following, we will first discuss why we consider $L$ parallel phase shifter networks and not any other arbitrary number such as 3$L$. 

\subsection{Why $L$ Parallel Phase Shifter Networks?}
Considering that ZF, i.e. pseudo-inverse operation, in fully-digital systems provides a near-optimal performance in rich scattering massive MIMO scenarios, we are aiming to design the hybrid beamformer according to the same principle. In other words, the first stage beamformer, i.e. RF beamformer, should maximize the SNR and the second-stage beamformer, i.e. baseband combiner, should mitigate the interference. From a signal processing perspective, performing the MF operation in the time or frequency domains result in the same performance for the fully-digital systems. However, performing MF in the frequency domain requires beamformer to be able to multiply each subcarrier of the signal with $\tilde{\mathbf{H}}^\text{H}(k)$ which is not feasible by the traditional hybrid beamformers. As discused, MF requires $L$ filter taps in the time-domain and this can be achieved by the $L$-tap structure in Fig. 1. This motivates us to investigate a scenario where there are $L$ parallel phase shifter networks to achieve a performance similar to MF. Then, we will evaluate 1-tap beamformer which is equivalent to the traditional hybrid beamforming.

\subsection{Proposed Method}
In the hybrid structures, it is not possible to directly apply MF due to the constant modulus constraint that is imposed by phase shifters. Considering MF as an $L$-tap filter, we design $\mathbf{W}_\text{RF}$ for the $L$-tap RF beamformer (please see Fig. 1) based on minimum mean square error (MMSE) criteria as
\begin{align}
\mathbf{W}_{\text{RF},l}^\star&= \underset{\mathbf{W}_{\text{RF},l}}{\arg\min}  \Vert \mathbf{W}_{\text{RF},l}-\mathbf{W}_{\text{MF},l} \Vert^2, \\ \nonumber
&= \underset{\mathbf{W}_{\text{RF},l}}{\arg\min} \frac{1}{\sqrt{M}} \sum_{m=1}^M \sum_{u=1}^U \vert \text{e}^{j\theta_{lum}} - H^\ast_{(-l)mu} \vert^2.
\end{align}
It could be easily verified that MMSE criteria is met when 
\begin{equation}
\label{eq:RFBeamformerL}
\mathbf{W}_\text{RF} (n)= \frac{1}{\sqrt{M}}\text{exp}\big( j\angle \mathbf{H}^\text{H}(-n) \big),
\end{equation}
or in other words $\theta_{lum}=-\angle H_{(-l)mu}$. Since this beamformer is in the form of equal gain combining, it also maximizes the SNR of the user signals. 

It is noted that $\lim_{M\to \infty}\mathbf{W}_{\text{RF},l}^\text{H}\mathbf{W}_{\text{RF},l'}=\mathbf{I}_K\delta(l-l')$ since the elements of the channel matrix are zero-mean i.i.d. random variables. As a result, the proposed RF beamformer does not result in noise coloring effect. Assuming that the first channel tap has the highest gain, the corresponding 1-tap RF beamformer that maximizes the received SNR is obtained by setting
\begin{equation}
\label{eq:RFBeamformer1}
\mathbf{W}_\text{RF}(n)=  \frac{1}{\sqrt{M}}\text{exp}\big( -j\angle \mathbf{H}_0^\text{H}\big)\delta (n).
\end{equation}
Using the same performance metrics as in \cite{6666103,6189014}, the RMS delay spread and the achievable sum-rate by the RF beamformers in (\ref{eq:RFBeamformerL}) and (\ref{eq:RFBeamformer1}) will be analyzed in the following. In addition, we will provide an asymptotic expression which provides a good approximation of the capacity of the effective channels by the 1 and $L$ tap beamformers.

\textbf{Proposition 1:} When the $L$-tap RF beamformer in (\ref{eq:RFBeamformerL}) is used the achievable sum-rate $R_{\text{sum}}^{L\text{-tap}}$, for $M\to \infty$, is given by $R_{\text{sum}}^{L\text{-tap}}= \sum_{u=1}^U \log_2 (1+ \gamma_k^{L\text{-tap}}),$ where the signal-to-interference-pluse-noise ratio (SINR) $\gamma_k^{L\text{-tap}}$ of the $k$-th user is
\begin{equation}
\label{eqSINR}
\gamma_k^{L\text{-tap}}=\frac{1 }{L\sigma_\text{z}^2+ P_\text{t}  (UL-1)} \times \frac{\pi P_\text{t}M }{4} \bigg \vert \sum_{l=0}^{L-1} d_{lu}^{1/2}\bigg \vert ^2.
\end{equation}

\textit{Proof:} Please refer to Appendix \ref{Prop1}. \hfill\(\Box\)

\textbf{Proposition 2:} When the $1$-tap RF beamformer in (\ref{eq:RFBeamformer1}) is used, and for $M\to \infty$, the achievable sum-rate $R_{\text{sum}}^{1\text{-tap}}$ is $R_{\text{sum}}^{1\text{-tap}}= \sum_{u=1}^U \log_2 (1+ \gamma_u^{1\text{-tap}}),$ where the SINR $\gamma_k^{1\text{-tap}}$ of the $u$-th user is
\begin{equation}
\gamma_k^{1\text{-tap}}=\frac{ d_{0u} }{\sigma_\text{z}^2+ P_\text{t} \sum_{l=1}^{L-1}    d_{nu} + P_\text{t}  (U-1)} \times \frac{\pi P_\text{t}M }{4}.
\end{equation}

\textit{Proof:} Please refer to Appendix \ref{Prop2}. \hfill\(\Box\)

Similar to fully-digital MF \cite{6189014}, propositions 1 and 2 indicate that the SNR by the 1-tap and $L$-tap beamformers increases proportional to $M$; however, achievable sum-rates reach a performance ceiling at high SNR regime.

\textbf{Proposition 3:} When the proposed RF beamformers in (\ref{eq:RFBeamformerL}) and (\ref{eq:RFBeamformer1}) are used, the RMS delay spread of each user's channel reduces with $1/\sqrt{M}$ when $M\to \infty$.

\textit{Proof:} Please refer to Appendix \ref{Prop3}. \hfill\(\Box\)

In terms of RMS delay spread proposition 3 indicates that the proposed RF beamformer presents a similar behavior as fully-digital MF in \cite{6666103}. Smaller RMS delay spread at the effective channel is equivalent to the statement that the effective channel tends to become frequency-flat. Motivated by this idea, proposition 4 exploits the deterministic behaviors of massive MIMO systems to derive the closed-form expression of the capacity of the effective channel.

\textbf{Proposition 4:} When the proposed RF beamformers in (\ref{eq:RFBeamformerL}) and (\ref{eq:RFBeamformer1}) are used and $M \to \infty $, the capacity of the effective channels by the $L$-tap and 1-tap beamformers are
\begin{align}
C(\mathbf{H}_\text{e,$L$-tap})= \log_2 \text{det} \big ( \mathbf{I}_U + \frac{\rho \pi M}{4} \big(  \sum_{l=0}^{L-1}\mathbf{D}_{l}^{1/2} \big)^2 \big ) ,
\end{align}
and
\begin{align}
C(\mathbf{H}_\text{e,$1$-tap})= \log_2 \text{det} \big ( \mathbf{I}_U + \frac{\rho \pi M}{4} \mathbf{D}_{0} \big ) ,
\end{align}
respectively.

\textit{Proof:} Please refer to Appendix \ref{Prop4}. \hfill\(\Box\)

In order to find a hybrid beamformer which is designed according to pseudo-inverse of the channel matrix, we apply ZF per subcarrier to mitigate the residual interference by the RF beamformer. It is noted that \textbf{x} can be created according to multicarrier or single-carrier techniques as the design of the RF beamformer is independent of the modulation of \textbf{x}. In other words, $\mathbf{W}_\text{RF, 1-tap}$ and $\mathbf{W}_\text{RF, $L$-tap}$ are solely designed according to the channel impulse response whereas the baseband combiner can be adjusted according to the modulation type of \textbf{x}, e.g. OFDM.  

\textit{Remark 1:} It is noted that our RF beamformer can be directly calculated from the phase of the elements of $\mathbf{H}(n) \in \mathbb{C}^{M \times U}$. Interestingly, as the received signals travel through the RF beamformer, equal gain combining is performed via RF phase shifters and adders. Hence, the RF beamformer of the proposed technique can reduce the digital signal processing that is needed at the baseband. Since digital ZF is performed over $\mathbf{H}_\text{e}(n)\in \mathbb{C}^{U \times U}$ and $K$ subcarriers, the complexity of proposed beamformer is related to $O(KU^3)$.

\subsection{How Many Parallel Phase Shifter Networks is Enough?}
After analyzing the performance of $L$-tap beamformer, the natural question is that how many parallel phase shifter networks are required to achieve the performance of a fully-digital beamforming with ZF per subcarrier? As discussed before, fully-digital beamforming with pseudo-inverse per subcarrier is equivalent to performing MF in the time domain followed by a matrix inversion per subcarrier. Hence, it could be easily verified that it suffices to achieve the performance of digital MF in the time domain with the RF beamformer. In this direction, first let's define 
\begin{equation}
\bar{\mathbf{W}}_{\text{MF},l}=\mathbf{W}_{\text{MF},l}/\gamma,
\end{equation}
where $\gamma=\max \: \vert W_{\text{MF},uml}\vert , \: \forall m \in \lbrace 1,...,M\rbrace, \: u \in \lbrace 1,...,U\rbrace, \: l \in \lbrace 1,...,L\rbrace$; and hence $\bar{H}_{\text{w},mul}\leq 1$. It could be easily verified that the normalization factor $\gamma $ does not have any impact on the system design and performance and both $\bar{\mathbf{W}}_{\text{MF},l}$ and $\mathbf{W}_{\text{MF},l}$ will result in the same result. On the other hand, for any arbitrary complex number $a$ where $0 \leq \vert a \vert \leq  1$ it can be concluded that 
\begin{align}
\vert a \vert & \text{e}^{j\angle a } =\text{e}^{j\angle a}\cos \Big(\cos^{-1} (\vert a \vert)\Big) \\ \nonumber 
&= \frac{\text{e}^{j\angle a }}{2}\text{e}^{j\cos^{-1} (\vert a \vert)}+\frac{\text{e}^{j\angle a }}{2}\text{e}^{-j\cos^{-1} (\vert a \vert)} \\ \nonumber
&=\frac{1}{2}\text{e}^{j\angle a+ j\cos^{-1} (\vert a \vert)}+\frac{1}{2}\text{e}^{j\angle a  - j\cos^{-1} (\vert a \vert) }.
\end{align}
This identity indicates that by having 2 parallel phase shifter networks per $\bar{\mathbf{W}}_{\text{MF},l}$, the RF beamformer will be able to fully-reconstruct $\bar{\mathbf{W}}_{\text{MF},l}$. In other words, by using $2L$ phase shifter networks, or equivalently two phase shifter per channel tap, the performance of a fully-digital ZF can be achieved. This is an extension to \cite{MOlischAntennaselection2005Journal,7542170} where the required number of the phase shifters was found to achieve the performance of fully-digital systems in frequency-flat channels.

\section{Simulation Results}
\label{sec:SecSR}
In this section, we use Monte Carlo simulations over 1000 channel realizations to evaluate the performance of the proposed methods and the closed-forms in Propositions 1 to 4. In addition to frequency-selective i.i.d. Rayleigh fading channel model, we will evaluate the performance over sparse scattering scenario. We will also provide performance comparisons with the 1-tap beamformer of \cite{7913599}. In the follwoing, let $\mathbf{H}_{\text{e},L\text{-tap}}$, $\mathbf{H}_{\text{e},1\text{-tap}}$, $\mathbf{H}_{\text{e,\cite{7913599}}}$ denote the effective channel matrix when the RF beamformers in (\ref{eq:RFBeamformerL}), (\ref{eq:RFBeamformer1}) and \cite{7913599} are applied, respectively. Moreover, $C(\mathbf{H}_{\text{e},i})$ and $R_\text{ZF}(\mathbf{H}_{\text{e},i})$, $i, \in \lbrace \text{MF, \textit{L}-tap, 1-tap, \cite{7913599}} \rbrace$, represent the capacity of $\mathbf{H}_{\text{e},i}$ and achievable sum-rate of $\mathbf{H}_{\text{e},i}$ when ZF is applied per subcarrier, respectively. In the following, the capacity of the effective channel $C(\mathbf{H}_{\text{e},i})$ is calculated by replacing $\mathbf{H}$ in (\ref{eq:fsCapacity}) with $\mathbf{H}_{\text{e},i}$. Unless otherwise stated, the simulation parameters are set as $M=100$, $U=4$, $K=128$, and the wireless channel is modeled by i.i.d. Rayleigh fading with $L=4$.

\begin{figure}
    \centering
          \includegraphics[width=.5\textwidth]{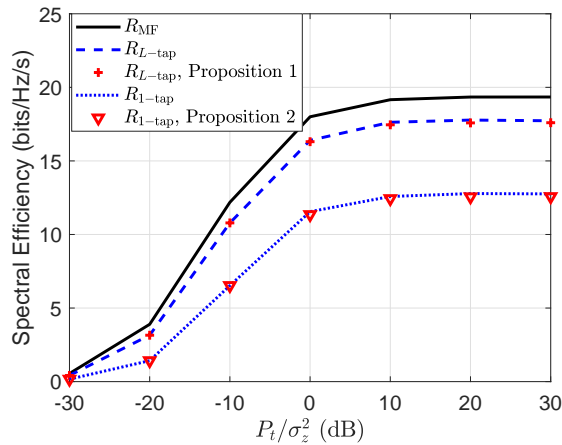}
        \caption{Spectral efficiency v.s. $P_\text{t}/\sigma_\text{z}^2$ for rich scattering and frequency-selective channel with $L=4$.}
        \label{fig:Single_carrier}
\end{figure}

\begin{figure}
    \centering
          \includegraphics[width=.5\textwidth]{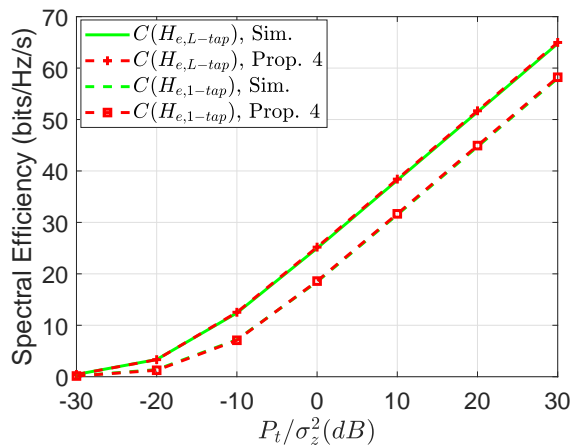}
        \caption{Capacity v.s. $P_\text{t}/\sigma_\text{z}^2$ by simulations and closed-forms of proposition 4 for rich scattering and frequency-selective channel with $L=4$.}
        \label{fig:Closed-forms}
\end{figure}

The closed-forms in propositions 1, 2 and 4 are shown in Fig. \ref{fig:Single_carrier} and Fig. \ref{fig:Closed-forms} and it is observed that there is a perfect match between propositions 1, 2, 4 and the simulations. Without baseband processing, Fig. \ref{fig:Single_carrier} indicates that the $L$-tap RF beamformer is capable of providing a similar performance compared to that of fully-digital MF in the low SNR regime. In the high SNR regime the performance of MF and the RF beamformers reach a performance ceiling. On the other hand, Fig. \ref{fig:Closed-forms} shows the capacity of the effective channels and illustrates the performance upper-bound by the baseband combiner; which indicates the best possible performance by a digital combiner.

\begin{figure}
    \centering
          \includegraphics[width=.5\textwidth]{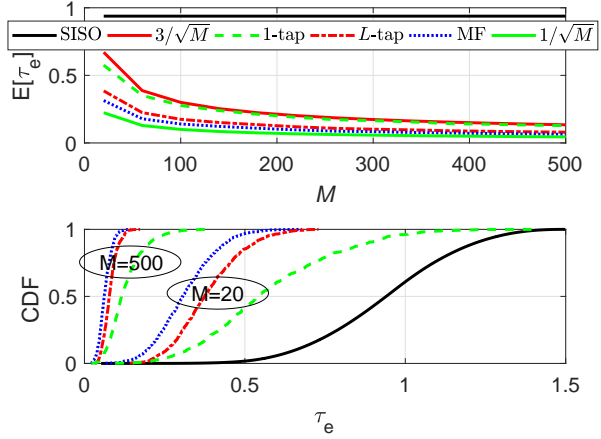}
        \caption{Top: Average RMS delay spread v.s. $M$. Bottom: CDF of RMS delay spread. }
        \label{fig:rms}
\end{figure}

\begin{figure}
    \centering
          \includegraphics[width=.49\textwidth]{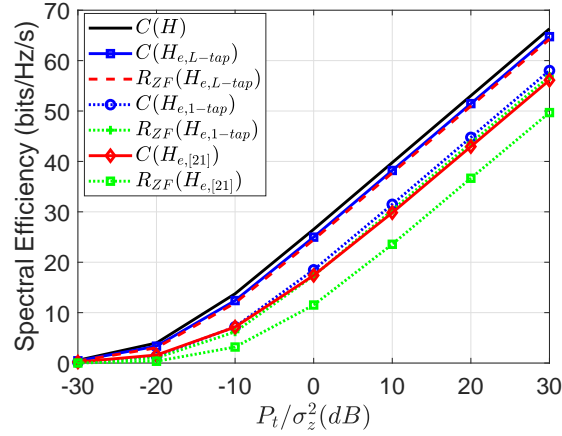}
        \caption{Spectral efficiency v.s. $P_\text{t}/\sigma_\text{z}^2$ for rich scattering and frequency-selective channel with $L=4$.}
        \label{fig:Performance1}
\end{figure}

\begin{figure}
    \centering
          \includegraphics[width=.49\textwidth]{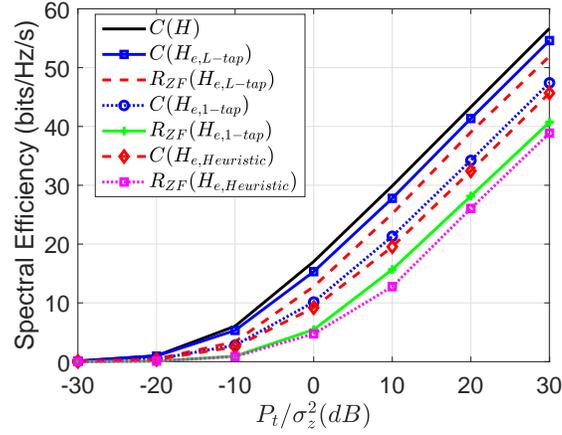}
        \caption{Spectral efficiency v.s. $P_\text{t}/\sigma_\text{z}^2$ for heuristic method over rich scattering and frequency-selective channel with $L=4$.}
        \label{fig:Reviewer1}
\end{figure}

Figure \ref{fig:rms} shows the average and cumulative distribution function (CDF) of the RMS delay of the effective channel of the users. As a comparison reference, we consider the RMS delay spread of the single-input single-output (SISO) channels between each user and each base station antenna. As discussed in Proposition 3, the average RMS delay spread of the effective channels by the proposed beamformers reduces with $1/\sqrt{M}$ as $M$ increases. In order to further clarify this behavior in Fig. \ref{fig:rms}, the average RMS delay spread curves always remain between the lower-bound $1/\sqrt{M}$ and upper-bound $3/\sqrt{M}$ curves where 1 and 3 are arbitrary coefficients to scale $1/\sqrt{M}$ and they are found via simulations. In terms of performance, MF and $L$-tap beamformers have similar average RMS delay spread which are lower than 1-tap. Moreover, CDF curves in Fig. \ref{fig:rms} shows that increasing the number of the antenna $M$ from 20 to 500 results in a significantly steeper curves resulting in a more stable and deterministic behavior for the RMS delay spreads observed at the baseband. 
Figure \ref{fig:Performance1} presents the performance of the proposed techniques in terms of capacity $C(\mathbf{H}_{\text{e},i})$ and the achievable sum-rates $R_\text{ZF}(\mathbf{H}_{\text{e},i})$ where $i, \in \lbrace \text{\textit{L}-tap, 1-tap, \cite{7913599}} \rbrace$. It is observed that $C(\mathbf{H}_{\text{e,}L\text{-tap}})$ and $R_\text{ZF}(\mathbf{H}_{\text{e},L\text{-tap}})$ are almost equal, and they are slightly lower than $C(\mathbf{H})$. Moreover, $C(\mathbf{H}_{\text{e,MF}})$ is almost the same as the capacity of the wireless channel $C(\mathbf{H})$. On the other hand, $C(\mathbf{H}_{\text{e},1\text{-tap}})$ and $R_\text{ZF}(\mathbf{H}_{\text{e},1\text{-tap}})$ by the 1-tap beamformer experience SNR losses compared to $L$-tap beamformer; however, same multiplexing gain are achieved with simpler circuitry compared to the $L$-tap beamformer. Figure \ref{fig:Performance1} also indicates that $C(\mathbf{H}_{\text{e,\cite{7913599}}})$ is lower than the $R_\text{ZF}(\mathbf{H}_{\text{e},1\text{-tap}})$ by our approach. In addition, $R_\text{ZF}(\mathbf{H}_{\text{e},1\text{-tap}})$ provides significantly higher spectral efficiency compared to $R_\text{ZF}(\mathbf{H}_{\text{e,\cite{7913599}}})$ when the RF beamformer of \cite{7913599} is combined with ZF per subcarrier at the baseband. 
Once may also consider a heuristic approach where the phase of the 1-tap beamformer is calculated according to the sum of the $L$ channel taps rather than using only the first tap. Figure. \ref{fig:Reviewer1} shows that both methods can achieve the same spatial multiplexing gain but the beamformer in equation (\ref{eq:RFBeamformer1}) results in higher capacity and achievable rate for the effective channel. This behavior is due to the fact that the 1-tap beamformer in (\ref{eq:RFBeamformer1}) can more efficiently collect the signal energy from the strongest channel tap and provide better SNR gains compared to the heuristic approach.  

\begin{figure}
    \centering
          \includegraphics[width=.49\textwidth]{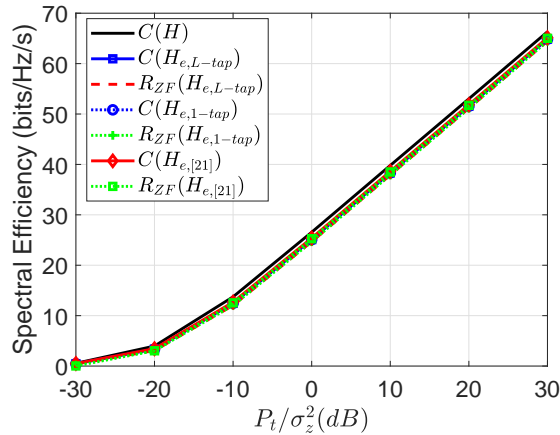}
        \caption{Spectral efficiency v.s. $P_\text{t}/\sigma_\text{z}^2$ for rich scattering and frequency-flat channel with $L=1$.}
        \label{fig:flat}
\end{figure}

It is noted that for $L=1$, i.e. frequency-flat channel, our proposed RF beamformers becomes the same as narrow-and beamformer in \cite{6928432,Globcome2018} and \cite{7913599} turns into \cite{7389996}. When $L=1$, Fig. \ref{fig:flat} shows that using the RF beamformer of \cite{7389996} results in a slightly higher spectral efficiency than \cite{6928432,Globcome2018}; however, this is achieved at the cost of higher complexity.

Although i.i.d. Rayleigh fading channel model is commonly used in the massive MIMO literature to present theoretical studies over the rich scattering channels \cite{marzetta_larsson_yang_ngo_2016}, this model is not a generic model for many practical scenarios \cite{6206345}. In order to investigate the performance of our proposed hybrid beamformer under more realistic channels, in the following, we consider a geometry-based channel model with uniform and linearly spaced antennas at the base station. As in \cite{7913599}, we assume that the channel between the base station and user $u$ consists of $L$ clusters such that $L$ taps in the time domain are observed. Moreover, the center of each cluster is uniformly distributed over $[0,2\pi)$, and the $N_\text{Sc}$ multipath components (MPCs) in each cluster follow Laplacian distribution with an angular spread of 10 degrees around the center of the cluster. In this model, the channel vector $\mathbf{h}^\text{T}_u \in \mathbb{C}^{N \times 1}$ for user $u$ is
\begin{equation}
\label{eq:SparseChannel}
\mathbf{h}_u=\sqrt{\dfrac{M }{LN_\text{Sc}}} \sum_{l=0}^{L-1} \sum_{i=1}^{N_\text{Sc}} \beta_{liu}\mathbf{a}(\phi_{liu})\delta(n-l),
\end{equation}
where $\beta_{liu}\sim \mathcal{CN}(0,d_{lu})$ is the multipath coefficient, $\phi_{liu}$ is the angle-of-arrival of the $i$th MPC in the $l$th cluster. The steering vector $\mathbf{a}(\phi_{liu})$ for linear arrays is 
\begin{align}
\label{Eq:Manifoldvector}
\mathbf{a}(\phi_{liu})=\dfrac{1}{\sqrt{M}}(1, \text{e}^{\frac{j2\pi d}{\lambda} \cos(\phi_{liu})}\: ...,\: \text{e}^{\frac{j2\pi d}{\lambda}(M-1)\cos(\phi_{liu})})^\text{T}
\end{align}
where $\phi_{liu}\in [0,\: \pi]$, $\lambda$ is the wavelength and $d=\lambda/2$ is the antenna spacing. The parameter $d_{lu}$ can be set according to different path loss models \cite{7913599}; however, without loss of generality and for the sake of consistency throughout the paper, we use (\ref{d_lk}) to set $d_{lu}$. It is noted that as this assumption does not impact our interpretation of the simulation results and performance evaluations.

\begin{figure}
    \centering
          \includegraphics[width=.49\textwidth]{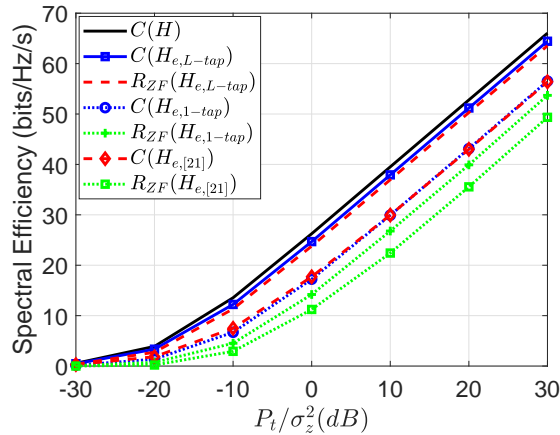}
        \caption{Spectral efficiency v.s. $P_\text{t}/\sigma_\text{z}^2 $ for sparse scattering and frequency-selective channel with $L=4$ and $N_\text{Sc}=5$.}
        \label{fig:HMShybridSparse_Selective}
\end{figure}

Similar to the rich scattering scenario, figures \ref{fig:HMShybridSparse_Selective} demonstrate the capacity $C(\mathbf{H}_{\text{e},i})$ and the achievable sum-rates $R_\text{ZF}(\mathbf{H}_{\text{e},i})$ over the sparse channel. For the frequency-selective scenario, with $N_\text{Sc}=5$ and $L=4$, it is observed that our proposed 1-tap RF beamformer and \cite{7913599} almost achieve the same capacity. However, the combination of our method and ZF per subcarrier results in a higher sum-rate compared to \cite{7913599}. 

\section{Conclusion}
In this paper the properties of the massive MIMO channels \textit{both in the time and frequency} domains have been exploited to design a low-complexity hybrid beamformer for frequency-selective channels. In the proposed approach, the RF beamformer is designed such that it coherently adds up the desired signals in the time domain. As a result, the effective channel has a much smaller RMS delay spread and its frequency response is more flat compared to the propagation channel. 
%In addition, the achievable sum-rate by the RF beamfomer provides a promising performance at the low-SNR regimes. However, the system performance saturates at high SNRs due to the residual interference. To mitigate this problem, the baseband combiner applies beamforming techniques in the frequency domain. The proposed beamformer provides a better performance compared to state-of-the-art in both rich and sparse scattering channels. 
The closed-form expressions derived in this work can also be used a design guide by the researchers to evaluate the performance of hybrid beamformers. By investigating new hybrid beamforming structures with larger number of phase shifters, it is shown that 2$L$ parallel networks are required to achieve the performance of a fully-digital ZF. Our proposed approach on designing hybrid beamformers for frequency-selective channels provides a fresh viewpoint to the problem and gives raise to new questions regarding further optimization of  $L$-tap beamformer for other channel models.

\appendices

\section{Proof of Poroposition 1}
\label{Prop1}
At $n=0$, the SINR for user $u$ is derived by
\begin{equation}
\gamma_u^{L\text{-tap}}=\frac{S_u}{\sigma_\text{z,e,$L$-tap}^2+I_\text{MUI}+
I_\text{ISI}},
\end{equation}
where $S_u$, $\sigma_\text{z,e,$L$-tap}^2$, $I _\text{MUI}$,
$I_\text{ISI}$ represent the power of the desired signal, noise, inter-user and inter-symbol interference for user $u$ at $n=0$. In order to calculate each of these parameters we firstly calculate the power delay profile (PDP) of the the effective channel. According to (\ref{ImpulseResponsesLtap}), the impulse response of the effective channel at the output of the $L$-tap beamformer is 
\begin{equation}
\mathbf{H}_\text{e, $L$-tap}(n)=\sum_{-L+1}^{L-1} \mathbf{W}_{\text{RF},{n-l}}  \mathbf{H}_l .
\end{equation}
Hence, the PDP of the effective channel for user $u$ is 
\begin{equation}
\label{eq:PDPcasesL}
P_{uu}(n)=
\begin{cases}
\bigg \vert \frac{\sum_{m=1}^M \sum_{l=0}^{L-1+n}  F_{lmun}}{\sqrt{M}} \bigg \vert^2, &  -L+1 \leq n <0,\\
\bigg \vert\frac{\sum_{m=1}^M \sum_{l=0}^{L-1}   \vert H_{lmu} \vert}{\sqrt{M}} \bigg \vert^2 ,& n=0,\\
\bigg \vert \frac{ \sum_{m=1}^M \sum_{l=n}^{L-1}  F_{lmun}}{\sqrt{M}} \bigg \vert^2 ,& 0<n \leq L-1 .\\
\end{cases}
\end{equation}
where $ F_{lmun}=H_{lmu} \text{e}^{-j\angle H_{(-\vert n \vert +l)mu}}$. In the following, the signal, noise and interference levels will be separately calculated. Since E$[\mathbf{x}(n) \mathbf{x}^\text{H}(n)]= P_\text{t} \mathbf{I}_U$, the expected value of the power $S_u$ from user $u$ at $n=0$ is 
\begin{align}
\label{eq:SignalPower}
S_u&=P_\text{t}P_{uu}(0)=P_\text{t}\bigg \vert\frac{\sum_{m=1}^M \sum_{l=0}^{L-1}   \vert H_{lmu} \vert}{\sqrt{M}} \bigg \vert^2\\ \nonumber
&=P_\text{t}M\bigg \vert\sum_{l=0}^{L-1} \frac{\sum_{m=1}^M  \vert H_{lmu} \vert}{M} \bigg \vert^2  \\ \nonumber
&= P_\text{t} M \bigg \vert  \sum_{l=0}^{L-1} \text{E}_M \big[ \vert H_{lmu} \vert \big] \bigg\vert^2 \\ \nonumber
&=P_\text{t}M\bigg \vert\sum_{l=0}^{L-1} d_{lk}^{1/2} \text{E}_M\big[  \vert H_{\text{w},lmu} \vert \big]\bigg \vert^2=\frac{\pi P_\text{t}M}{4}\bigg \vert \sum_{l=0}^{L-1} d_{lu}^{1/2}\bigg \vert ^2,
\end{align}
as $\text{E}_M\big[  \vert H_{\text{w},lmu} \vert \big]=\sqrt{\pi}/2$ due to the Gaussian distribution of its elements \cite{7542170}. To calculate the intersymbol interference, it could be easily verified that 
\begin{equation} 
I_{\text{ISI},u}=P_\text{t}\big[\sum_{ n=-L+1}^{-1} P_{uu}(n)+ \sum_{ n= 1}^{L-1} P_{uu}(n)\big].
\end{equation} 
To analyze the first term, let us define random variable $G_m= \sum_{l=0}^{L-1+n} F_{lmun}$ where $\text{E}_M[G_m]=0$ and $n<0$. In addition, $G_m$ and $G_{m'}, \forall m\neq m'$ are independent and uncorrelated as $H_{lmu}$ and $H_{lm'u}$ are independent and uncorrelated. Hence, $\text{E}_M[G_m G_{m'}]=\text{E}_M[G_m] \text{E}_M[G_{m'}]=0$. Applying the law of large numbers to (\ref{eq:PDPcasesL}), when $M \to \infty$ and $-L+1 \leq n <0$, leads to
\begin{align}
\label{eq:PDPOff}
&P_{uu}(n)=  \bigg\vert  \frac{1} {\sqrt{M}} \sum_{m=1}^M G_m \bigg\vert^2=\\ \nonumber
&= \frac{1}{M} \sum_{m=1}^M \vert g_m \vert^2+  \frac{1}{M}\sum_{m=1}^M \sum_{\substack{m'=1 \\ m' \neq m}}^M G_m G_{m'} \\ \nonumber
& = \text{E}_M\big [ \vert  G_m \vert^2 \big] +\text{E}_M \big[  G_m G_{m'} \vert\big]= \text{E}_M\big [ \vert  G_m \vert^2 \big]\\ \nonumber
& \stackrel{(a)}{=} \text{Var}(G_m) =  \text{Var}(\sum_{l=0}^{L-1+n} F_{lmun})   \\ \nonumber
&\stackrel{(c)}{=}   \sum_{l=0}^{L-1+n} \text{Var}_M\big(F_{lmun} \big) =\sum_{l=0}^{L-1+n} d_{lk},
\end{align}
where (a) holds because $G_m$ and $G_{m'}$ are independent zero-mean random variables; and (b) is directly deduced from the definition of variance for random variable $a$ as $\text{Var}(a) = \text{E}[\vert a \vert^2]-\vert \text{E}[ a ]\vert^2 =\text{E}[\vert a \vert^2]$ when $\text{E}[ a ]=0$. Finally, (c) holds because for independent random variables $a$ and $b$, Var($a+b$)=Var($a$)+Var($b$). Similarly, $\forall 0<n\leq L-1$, it could be shown that $P_{kk}(n) =\sum_{l=n}^{L-1} d_{lk}$; hence,
\begin{align}
\label{eq:ISICalculation}
I_{\text{ISI},u}&= P_\text{t}\bigg( \sum_{n=-L+1}^{-1}   \sum_{l=0}^{L-1+n} d_{lu} +\sum_{n=1}^{L-1}   \sum_{l=n}^{L-1} d_{lu} \bigg)  \\ \nonumber
&=P_\text{t}\bigg(  \sum_{n=1}^{L-1}   \sum_{l=0}^{n-1} d_{lu} + \sum_{n=1}^{L-1}   \sum_{l=n}^{L-1} d_{lu}  \bigg)  \\ \nonumber
&= P_\text{t}\bigg( \sum_{n=1}^{L-1}   \sum_{l=0}^{L-1} d_{lu} \bigg) =(L-1)P_\text{t},
\end{align}
where the last equality comes from the normalization $\sum_{l=0}^{L-1} d_{lk}=1$. 

Moreover, the multiuser interference is
\begin{equation}
I_{\text{MUI},u}=P_\text{t}\sum_{ u\neq u'} \sum_{n}P_{uu'}(n), 
\end{equation}
$\forall u,u' \in \lbrace 1,...,U \rbrace$ and $u\neq u'$. Defining $F'_{lmkn}= \text{e}^{-j\angle H_{(-\vert n \vert +l)mu}} H_{lmu'}$, it is an i.i.d. zero-mean random variable with respect to $m$. For $ n \in \lbrace -L+1,...,-1\rbrace$, $P_{uu'}(n)$ is equal to
\begin{align}
&\bigg \vert \frac{1}{\sqrt{M}}\sum_{m=0}^{M} \sum_{l=0}^{L-1+n}  F'_{lmun} \bigg \vert^2 = \sum_{l=0}^{L-1+n} \text{E}_M\big [ \vert F'_{lmun} \vert^2 \big]\\ \nonumber
&=  \sum_{l=0}^{L-1+n} \text{Var}_M \big(F'_{lmun} \big)=\sum_{l=0}^{L-1+ n } d_{lu}.
\end{align}
It could be easily verified that $P_{uu'}(0)=1$ due to the normalization $\sum_{l=0}^{L-1} d_{lu}=1$, and $P_{uu'}(n > 0) = \sum_{l=n}^{L-1} d_{lu}$. Similar to (\ref{eq:ISICalculation}), the multiuser interference term $ I_\text{MUI}$ is related to
\begin{align}
\sum_{n=-L+1}^{L-1} P_{uu'}(n) &=P_{uu'}(0)+\sum_{n=-L+1}^{-1} P_{uu'}(n)  +\sum_{n=1}^{L-1} P_{uu'}(n)\\ \nonumber 
&=1+\bigg( \sum_{n=1}^{L-1}   \sum_{l=0}^{L-1} d_{lu} \bigg)=L.
\end{align}
Due to the symmetry of the problem, the total interference from $U-1$ users on user $u$ is $I_{\text{MUI}}= (U-1)LP_\text{t}.$ Since noise is a zero-mean i.i.d. random variable, the power $\vert z_\text{e}(0)\vert^2$ and variance of the effective noise $\sigma_\text{z,e,$L$-tap}^2$ at the baseband are equal. At $n=0$, $\vert z_\text{e}(0)\vert^2$ is  
\begin{align}
 \bigg \vert \sum_{l=0}^{L-1} \sum_{m=1}^{M}\frac{1}{\sqrt{M}} \text{e}^{-j\angle H_{lmu}}{z}_{ml}  \bigg \vert^2 = \sigma_\text{z}^2   \sum_{l=0}^{L-1} \text{E}_M [\vert{z}_{ml} \vert^2] = L\sigma_\text{z}^2 .
 \end{align} 
Finally, the SINR for use $u$ is
\begin{align}
\gamma_u^{L\text{-tap}}&=\frac{S_u}{\vert z_\text{e,$L$-tap}(0)\vert^2+I_\text{MUI}+
I_\text{ISI}}\\ \nonumber
&= \frac{\pi P_\text{t}M\bigg \vert \sum_{l=0}^{L-1} d_{lu}^{1/2}\bigg \vert ^2 /4 }{L\sigma_\text{z}^2+(L-1)P_\text{t} +(U-1)LP_\text{t}}.
\end{align}

\hfill\(\Box\)

\section{Proof of Poroposition 2}
\label{Prop2}
Similar to the proof of propositon 1, the PDP of the impulse response in (\ref{ImpulseResponses1tap}) will be used to calculate the desired signal, interference and noise power at the output of the RF beamformer. The impulse response of the effective channel at the output of the $1$-tap beamformer is 
\begin{equation}
\mathbf{H}_\text{e, $1$-tap}(n) = \mathbf{W}_{\text{RF}}\mathbf{H}(n) .
\end{equation}
Similar to Appendix \ref{Prop1}, the PDP of the effective channel for user $u$ is expressed as 
\begin{align}
P_{uu}(n) = \bigg \vert\frac{1}{\sqrt{M}} \sum_{m=1}^M  H_{nmu} \text{e}^{-j h_{0mu}}  \bigg \vert^2
\end{align}
$\forall n\in \lbrace 0,...,L-1 \rbrace$, otherwise $P_{uu}(n) =0$. Hence, the power of the desired signal for user $u$ is 
\begin{align}
S_u=P_\text{t}  P_{uu}(0) =  \bigg \vert\frac{1}{\sqrt{M}} \sum_{m=1}^M  \vert H_{0mu} \vert \bigg \vert^2=\frac{P_\text{t} \pi M d_{0u}}{4}.
\end{align}
Inter-symbol interference is
\begin{align}
 I_\text{ISI}=P_\text{t} \sum_{ n\neq 0} P_{uu}(n)=P_\text{t} \sum_{l=1}^{L-1}    d_{lu}.
\end{align}
Since the channels for users $u$ and $u'$ are independent and uncorrelated, the interference $I_{\text{MUI},u'}$ form user $u'$ is related to
\begin{align}
 &\sum_{n=0}^{L-1} P_{uu'}(n)=\sum_{n=0}^{L-1}\frac{1}{M}  \bigg\vert\sum_{m=0}^{M}  H_{lmu'} \text{e}^{-j\angle H_{0mu}} \bigg \vert^2 \\ \nonumber
 &=  \sum_{l=0}^{L-1} \text{Var}_M \big(H_{lmu'}  \big)=\sum_{l=0}^{L-1}  d_{lu'}
= 1.
\end{align}
Finally, the effective noise power at $n=0$ is $\vert z_\text{e}(n)\vert^2=\vert 1/\sqrt{M} \sum_{m=1}^{M} {z}_{ml}\vert^2 = \sigma_\text{z}^2 $, and the proposition can be easily proved. \hfill\(\Box\)

\section{Proof of Poroposition 3}
\label{Prop3}
In this appendix, we only present the proof for the $L$-tap beamformer as the same steps could be repeated for the 1-tap scenario. PDP of the effective channel of user $k$ was derived in (\ref{eq:PDPcasesL}). Moreover, the RMS delay spread $\tau_{\text{e},u} $ is 
\begin{equation}
\label{eq:rms}
\tau_{\text{e},u}=\sqrt{\frac{\sum_{n=-L+1}^{L-1}  P_{uu}(n) n^2 }{\sum_{n=-L+1}^{L-1}  P_{uu}(n)}-\bigg(\frac{\sum_{n=-L+1}^{L-1} P_{uu}(n) n }{\sum_{n=-L+1}^{L-1}  P_{uu}(n)} \bigg)^2}.
\end{equation}
It is noted that $P_{uu}(0)$ does not impact the numerators in (\ref{eq:rms}); however, it is included in the denominators. Moreover, according to Appendix \ref{Prop1}, $P_{uu}(0)$ increases with $M$ whereas $P_{uu}(n \neq 0)$ shows a deterministic behavior. As the ratio of the of the numerator to the denominator of (\ref{eq:rms}) is related to $1/M$, the RMS delay spread $\tau_\text{e}$ reduces with $1/\sqrt{M}$. \hfill\(\Box\)

\section{Proof of Proposition 4}
\label{Prop4}
Using (\ref{FFTeq}), the frequency response of the effective channels in (\ref{ImpulseResponsesLtap}) and (\ref{ImpulseResponses1tap}) are
\begin{align}
\label{freqHLtap}
\tilde{\mathbf{H}}_\text{e,$L$-tap}(k)&=\mathbf{H}_{\text{e,$L$-tap},0} +\sum_{\substack{l=-L+1 \\ l\neq 0}}^{L-1} \mathbf{H}_{\text{e, $L$-tap,}l} \text{exp}(-\frac{j2\pi l k}{K}),
\end{align}
and 
\begin{align}
\label{freqH1tap}
\tilde{\mathbf{H}}_\text{e, $1$-tap}(k)&=\mathbf{H}_{\text{e, $1$-tap},0} +\sum_{l=1}^{L-1} \mathbf{H}_{\text{e, $1$-tap,}l} \text{exp}(-\frac{j2\pi l k}{K}),
\end{align}
respectively. Law of large numbers indicate that when $M \to \infty$, the first term in (\ref{freqHLtap}) can be expressed as
\begin{align}
\mathbf{H}_{\text{e,$L$-tap},0} &= \sum_{l=0}^L \mathbf{W}_{\text{RF}, l}\mathbf{H}_{l}\\ \nonumber &=\sum_{l=0}^L\sqrt{M}\bigg[ \frac{\text{exp}\big( -j\angle \mathbf{H}_l^\text{H})\mathbf{H}_{\text{w},l} \mathbf{D}_{l}^{1/2}}   {M}\bigg] \\ \nonumber
 &\to \frac{\sqrt{\pi M}}{2} \sum_{l=0}^L\mathbf{D}_{l}^{1/2}.
\end{align}
Similarly, the first term in (\ref{freqH1tap}) is
\begin{align}
 \mathbf{H}_{\text{e,1-tap},0} &= \mathbf{W}_{\text{RF}, 0}\mathbf{H}_{0} \\ \nonumber
 &=\sqrt{M}\bigg[ \frac{\text{exp}\big( -j\angle \mathbf{H}_0^\text{H})\mathbf{H}_{\text{w},0} \mathbf{D}_{0}^{1/2}}   {M}\bigg]\to \frac{\sqrt{\pi M}}{2} \mathbf{D}_{0}^{1/2}.
\end{align}
Due to the zero-mean and uncorrelated property of the elements of $\mathbf{H}_\text{w,l}$, it could be easily verified that $\forall l\neq l'$
\begin{align}
 \frac{1}{\sqrt{M}}\mathbf{W}_{l'}\mathbf{H}_{l}\to \mathbf{0}_{U\times U}.
\end{align}
Hence, when $M \to \infty$ the frequency response of the effective channel by the $L$-tap and 1-tap beamformers are  
\begin{align}
\label{HFlatEquavalentLTAP}
\tilde{\mathbf{H}}_\text{e,$L$-tap}(k)= \frac{\sqrt{\pi M}}{2} \sum_{l=0}^L\mathbf{D}_{l}^{1/2},
\end{align}
and 
\begin{align}
\label{HFlatEquavalent}
\tilde{\mathbf{H}}_\text{e,1-tap}(k) = \frac{\sqrt{\pi M}}{2} \mathbf{D}_{0}^{1/2},
\end{align}
respectively. This indicates that $\tilde{\mathbf{H}}_\text{e,$L$-tap}(k)$ and $\tilde{\mathbf{H}}_\text{e,$1$-tap}(k)$ can be treated as a frequency-flat channel given by equations using equations (\ref{HFlatEquavalentLTAP}) and (\ref{HFlatEquavalent}), respectively. Finally, using (\ref{eq:fsCapacity}), the capacity of the effective channels by the $L$-tap and 1-tap beamformers in the limit of $M \to \infty$ (\ref{eq:fsCapacity}) become
\begin{align}
C(\mathbf{H}_\text{e,$L$-tap})&= \log_2 \text{det} \big ( \mathbf{I}_U + \rho  \mathbf{H}_{\text{e},0} \mathbf{H}_{\text{e},0}^\text{H}\big ) \\ \nonumber
&= \log_2 \text{det} \big ( \mathbf{I}_U + \frac{\rho \pi M}{4} \big(  \sum_{l=0}^L\mathbf{D}_{l}^{1/2} \big)^2 \big ) ,
\end{align}
and
\begin{align}
C(\mathbf{H}_\text{e,$1$-tap})&= \log_2 \text{det} \big \vert \mathbf{I}_U + \frac{\rho \pi M}{4} \mathbf{D}_{0} \big \vert ,
\end{align}
respectively.

\bibliographystyle{IEEEtran}


\end{document}